# MOONS: a Multi-Object Optical and Near-infrared Spectrograph for the VLT

M. Cirasuolo[1,2*], J. Afonso[3], R. Bender[4,5], P. Bonifacio[6], C. Evans[1], L. Kaper[7], E. Oliva[8], L. Vanzi[9], M. Abreu[10], E. Atad-Ettedgui[1], C. Babusiaux[6], F. Bauer[9], P. Best[2], N. Bezawada[1], I. Bryson[1], A. Cabral[10], K. Caputi[2], M. Centrone[15], F. Chemla[6], A. Cimatti[11], M-R. Cioni[12], G. Clementini[20], J. Coelho[10], E. Daddi[13], J. Dunlop[2], S. Feltzing[14], A. Ferguson[2], H. Flores[6], A. Fontana[15], J. Fynbo[16], B. Garilli[23], A. Glauser[17], I. Guinouard[6], F. Hammer[6], P. Hastings[1], A. Hess[4], R. Ivison[1], P. Jagourel[6], M. Jarvis[12], G. Kauffman[18], A. Lawrence[2], D. Lee[1], G. Licausi[15], S. Lilly[17], D. Lorenzetti[15], R. Maiolino[15], F. Mannucci[8], R. McLure[2], D. Minniti[9], D. Montgomery[1], B. Muschielok[4], K. Nandra[5], R. Navarro[19], P. Norberg[2], L. Origlia[20], N. Padilla[9], J. Peacock[2], F. Pedicini[15], L. Pentericci[15], J. Pragt[19], M. Puech[6], S. Randich[8], A. Renzini[21], N. Ryde[14], M. Rodrigues[24], F. Royer[6], R. Saglia[4,5], A. Sanchez[5], H. Schnetler[1], D. Sobral[2], R. Speziali[15], S. Todd[1], E. Tolstoy[22], M. Torres[9], L. Venema[22], F. Vitali[15], M. Wegner[4], M. Wells[1], V. Wild[2], G. Wright[1]

[1]STFC UK Astronomy Technology Centre, Edinburgh, UK; [2]Institute for Astronomy, Edinburgh, UK; [3]Observatorio Astronomico de Lisboa, Portugal; [4]Universitaets-Sternwarte, Munchen, Germany; [5]Max-Planck-Institut fuer Extraterrestrische Physik, Munchen, Germany; [6]GEPI, Observatoire de Paris, CNRS, Univ. Paris Diderot, France; [7]Astronomical Institute Anton Pannekoer, Amsterdam, The Netherlands; [8]INAF-Osservatorio Astrofisico di Arcetri, Italy; [9]Centre for Astro-Engineering at Universidad Catolica, Santiago, Chile, [10]Centre for Astronomy and Astrophysics of University of Lisboa, Portugal; [11]Università di Bologna - Dipartimento di Astronomia, Italy; [12]University of Hertfordshire, UK; [13]CEA-Saclay, France; [14]Lund Observatory, Sweden; [15]INAF-Osservatorio Astronomico Roma, Italy; [16]Dark Cosmology Centre, Copenhagen, Denmark; [17]ETH Zürich, Switzerland; [18]Max-Planck-Institut für Astrophysik, Garching, Germany; [19]NOVA-ASTRON, The Netherlands; [20]INAF-Osservatorio Astronomico Bologna, Italy; [21]INAF-Osservatorio Astronoomico Padova, Italy; [22]Kapteyn Astronomical Institute, Groningen, The Netherlands; [23]IASF-INAF, Milano, Italy; [24] European Southern Observatory, Santiago, Chile.

## ABSTRACT

*MOONS* is a new conceptual design for a **M**ulti-**O**bject **O**ptical and **N**ear-infrared **S**pectrograph for the Very Large Telescope (VLT), selected by ESO for a Phase A study. The baseline design consists of ~1000 fibers deployable over a field of view of ~500 square arcmin, the largest patrol field offered by the Nasmyth focus at the VLT. The total wavelength coverage is 0.8μm-1.8μm and two resolution modes: medium resolution and high resolution. In the medium resolution mode (R~4,000-6,000) the entire wavelength range 0.8μm-1.8μm is observed simultaneously, while the high resolution mode covers simultaneously three selected spectral regions: one around the CaII triplet (at R~8,000) to measure radial velocities, and two regions at R~20,000 one in the J-band and one in the H-band, for detailed measurements of chemical abundances.

The grasp of the 8.2m Very Large Telescope (VLT) combined with the large multiplex and wavelength coverage of *MOONS* – extending into the near-IR – will provide the observational power necessary to study galaxy formation and evolution over the entire history of the Universe, from our Milky Way, through the redshift desert and up to the epoch of re-ionization at z>8-9. At the same time, the high spectral resolution mode will allow astronomers to study chemical abundances of stars in our Galaxy, in particular in the highly obscured regions of the Bulge, and provide the necessary follow-up of the Gaia mission. Such characteristics and versatility make *MOONS* the long-awaited workhorse near-IR MOS for the VLT, which will perfectly complement optical spectroscopy performed by FLAMES and VIMOS.

---

[*]Further author information: Send correspondence to Michele Cirasuolo, email: ciras@roe.ac.uk

## 1. INTRODUCTION

In recent years, several large spectroscopic surveys at optical wavelengths (0.3μm - 1μm) have been undertaken and have provided key information on the formation and evolution of galaxies in the local Universe and up to z≈1 (when the Universe was about half of his current age). However, they have arguably now reached their limits and spectroscopy at λ>1μm is now crucial to extend our knowledge beyond z≈1, encompassing the peak of mass assembly and star-formation (1<z<3) and into the uncharted epochs at z>7. In fact, observations in the near-IR are essential because many of the objects of interest are red and therefore brighter in the near-IR compared to the optical, due to either **i)** extreme redshift, in the case of galaxies and black holes at z>7 or **ii)** dust obscuration, in the case of stars in the Bulge of our Galaxy and the extreme dust-enshrouded star-forming galaxies revealed by Herschel, or **iii)** age, in the case of the oldest, passively evolving galaxy population or **iv)** low intrinsic temperature, as in the case of low-mass stars.

To meet the aspirations of both the extragalactic and galactic scientific communities, we have presented to ESO a new conceptual design for a highly multiplexed spectrograph, *MOONS,* capable of combining both optical and near-IR capabilities, as well as both medium and high-resolution modes (Cirasuolo et al. 2011). *MOONS* was selected in conjunction with *4MOST* (see de Jong 2012, these proceedings) for a Phase A study, which will be completed by February 2013 and the decision on which project(s) will go ahead expected in Spring 2013 (Ramsay et al. 2011). The two concepts – *MOONS* and *4MOST* - go on different telescopes and therefore it is technically feasible to have both projects go forward.

This paper is organized as follows: in the next section we provide an overview of the main instrument science cases, in the next section we summarise the top level requirements, before providing in the subsequent section a brief overview of the instrument baseline design.

## 2. THE MAIN SCIENCE CASES

*MOONS* will be a versatile, world-leading instrument that will provide ESO and the astronomical community with the observational power necessary to tackle a wide range of Galactic, Extragalactic and Cosmological studies. Here we briefly highlight some of the main science cases that are driving the *MOONS* design.

### 2.1. Galactic Achaeology

The study of resolved stellar populations of the Milky Way and other Local Group galaxies can provide us with a fossil record of their chemo-dynamical and star-formation histories over many gigayear timescales. Scheduled for launch in 2013, the ESA Gaia mission will deliver new insight into the assembly history of the Milky Way, but to exploit its full potential ground-based follow-up is required. *MOONS* will provide this crucial follow-up for Gaia and for other ground based surveys with VISTA, Pan-STARRS, UKIDSS, by measuring accurate radial velocities, metallicities and chemical abundances for several millions of stars. Given the spectral resolutions (R~5000 and R~20,000) and the ability of observing in the near-IR, *MOONS* will perfectly complement the ongoing and planned surveys (see Figure 1) including the new large Gaia-ESO public spectroscopic survey. The unique features of *MOONS* will allow us in particular to clarify the nature of the extincted regions of the Bulge, but also to assess the chemo-dynamical structure of the Thin and Thick Disc, understand the importance of satellites and streams in the Halo, ultimately creating an accurate 3D map of our Galaxy to provide essential insight into its origin and evolution.

### 2.2. The growth of Galaxies

Tracing the assembly history of galaxies over cosmic time remains a primary goal for observational and theoretical studies of the Universe. Even though, in recent years, large spectroscopic surveys at optical wavelengths (0.3μm - 1μm) have provided key information on the formation and evolution of galaxies, near-IR spectroscopy is now crucial to extend our knowledge beyond z≈1. In fact, at these redshifts almost all the main spectral features are shifted at λ>1μm. Exploiting the large multiplex and wavelength coverage of *MOONS* it will be possible to create the equivalent of the successful Sloan Digital Sky Survey, but at z>1 (see Figure 2). This will provide an unparalled resource to study the physical processes that shape galaxy evolution and determine the key relations between stellar mass, star-formation, metallicity and the role of feedback. Filling a critical gap in discovery space, *MOONS* will be a powerful instrument to unveil ``the redshift desert'' (1.5<z<3, see Figure 2) and study this crucial epoch around the peak of star-formation, the assembly of the most massive galaxies, the effect of the environment and the connection with the shining of powerful active nuclei.

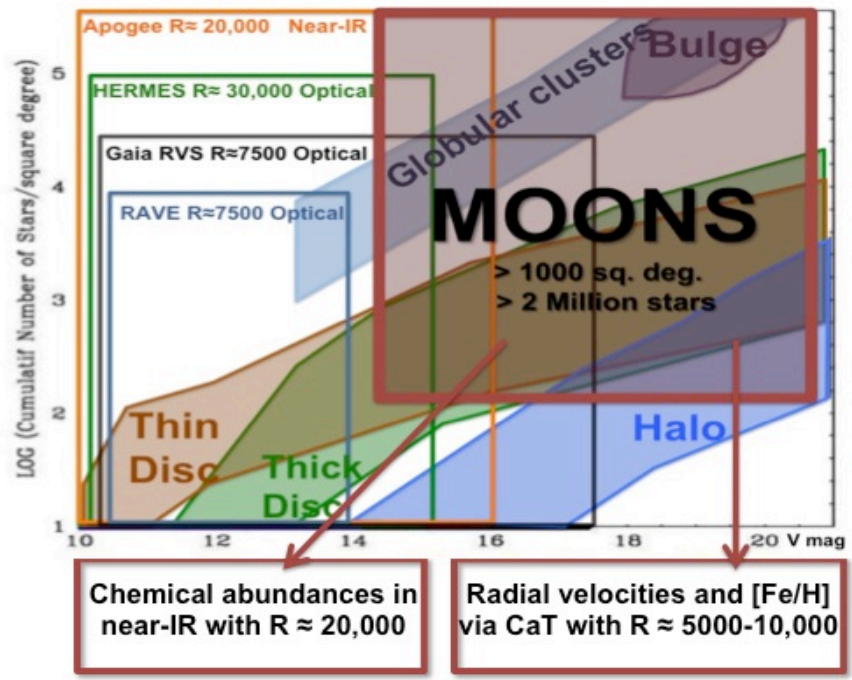

**Figure 1.** Number density of stars in the various components of the Milky Way as a function of V-band magnitude (figure adapted from Recio-Blanco, Hill, Bienaymé 2009). Gaia will provide astrometry for all stars with V<20, however the on-board spectrometer (RVS) will deliver chemical abundance only for stars brighter than 13$^{th}$ and radial velocities for stars brighter than 17$^{th}$ magnitude. *MOONS* will perfectly complement Gaia and the other spectroscopic surveys (e.g. Apogee, Hermes, RAVE) providing chemical abundances via high-resolution spectroscopy in the near-IR (e.g. observing Ca, Si, S, Fe, Ti lines) and radial velocities via Calcium triplet.

## 2.3. The first galaxies

The shining of the first galaxies, just few hundred million years after the Big Bang (at redshift 7<z<12) is of enormous importance in the history of the Universe since they hold the key to furthering our understanding of the cosmic re-ionization. Although recent advances obtained by deep near-IR imaging have been dramatic, very little is known about when and especially how this re-ionization happened. The unique combination of 8m-aperture, wide area coverage and near-IR spectroscopy (key since at z>7 even Lyman-α is shifted at λ>1μm) offered by *MOONS,* will provide accurate distances, relative velocities and emission line diagnostics, without which the power of these photometric surveys is severely limited. The capabilities of *MOONS* will give us the first realistic chance to perform a systematic, wide-area spectroscopic study of the very high redshift galaxies and establish the physics of re-ionization.

## 2.4. Cosmology

Over the last two decades several observational keystones have considerably changed our knowledge of the Universe. Measurements of the Cosmic Microwave Background, high-redshift super-novae and large-scale structure have revealed that 96% of the density of the Universe consists of currently unexplained Dark Energy and Dark Matter, and less than 4% is in the form of baryons. Understanding the nature of these dark components - which dominate the global expansion and large-scale structure – is amongst the most fundamental unsolved problems in science. Complementary to other spectroscopic surveys at z<1 (e.g. Vipers, BOSS, WiggleZ, BigBOSS), the capabilities of *MOONS* will allow us to constrain the cosmological paradigm of the Λ Cold Dark Matter model by determining the Dark Matter halo mass function and obtain crucial constraints on the nature of Dark Energy and gravity via detailed measurements of the growth rate of structure at z>1.

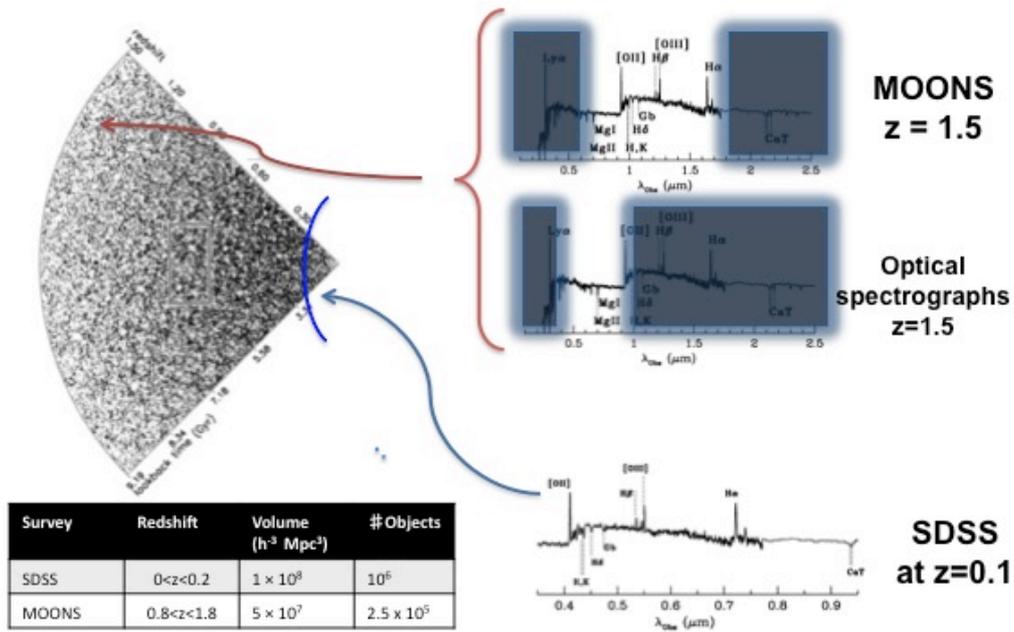

**Figure 2.** A medium-deep survey by *MOONS* at z>1 will provide a large number of spectra of the same quality and over the same rest-frame wavelength range and comoving volume as the low-redshift SDSS survey. The crucial redshift range 1.5<z<2.5, encompassing the peak of star-formation, has proved to be the hardest to explore spectrally (because the major features are redshifted out of the optical range) and gained the nickname of "redshift desert". *MOONS* will cover this gap and properly trace the evolution of galaxies throughout the redshift desert.

## 2.5. Synergy with large area imaging surveys and Euclid

*MOONS* is the ideal instrument to provide the essential deep spectroscopic follow up of imaging surveys undertaken with facilities in optical and near-IR (VISTA, UKIDSS, VST, Pan-STARRS, Dark Energy Survey, LSST) and facilities operating at other wavelengths (ALMA, Herschel, eRosita, LOFAR, WISE, ASKAP).

*MOONS* will play also an important role for the recently approved ESA mission Euclid, which will make use of near-IR slitless spectroscopy over very wide sky areas, covering the 0.8-1.8μm spectral range in order to detect emission line galaxies (e.g. Hα emitters at 0.5<z<2). It is well known that slitless spectroscopy is affected by "confusion" problems due to the overlap of spectra coming from different objects. By observing specific selected fields, *MOONS* will cover the same spectral range of the space observations and provide uncontaminated and higher resolution spectra of galaxies that will be very useful to independently assess the level and impact of spectral "confusion", as well as to investigate and correct potential biases introduced by slitless spectroscopy (e.g. redshift accuracy, success rate, emission line properties, galaxy types, etc).

## 2.6. The consortium.

Reflecting somehow the wide range of science goals, the *MOONS* consortium is built on the expertise of several partners: Chile, Denmark, France, Germany, ESO, Italy, Netherlands, Portugal, Sweden, Switzerland and United Kingdom.

# 3. SCIENCE REQUIREMENTS

The essential requirements derived from the science cases indicate the necessity of a multiplex of ~1000 objects to be observed at wavelengths between 0.8μm to1.8μm, with the possibility to select two resolution modes: a resolving power of R > 4,000 and a high resolution mode at R≈20000. A summary of the main requirements is presented in Table 1.

| Parameter | Requirement |
| --- | --- |
| Telescope | VLT |
| Field of View | 500 square arcmin |
| Multiplex | 1000 fibers, with possibility to deploy in pairs (500 obj+500 sky) |
| Sky-projected diameter of single fiber | 1.2 arcsec |
| Wavelength coverage | 0.8μm-1.8μm |
| Observing mode | medium resolution (MR) and high resolution (HR) |
| Simultaneous λ-coverage in MR | 0.8μm-1.8μm |
| Simultaneous λ-coverage in HR | [0.8μm-0.9μm] + [1.17μm-1.26μm] + [1.52μm-1.63μm] |
| Resolving power in MR | R~ 4,000 – 6,000 |
| Resolving power in HR | R≥7,000 (around CaII triplet); R≥20,000 (in the J and H bands) |

**Table 1.** Instrument requirements for MOONS baseline.

## 3.1. Sky subtraction

An accurate subtraction of the sky background is critical when observing faint sources, particularly in the near-infrared, where the background is dominated by strong OH sky lines. To achieve this goal, a two-fold approach has been foreseen for *MOONS*: 1) high spectral resolving power and 2) ability to perform observations in cross-beam switching mode.

The high resolving power, even in the Medium Resolution mode (R>4000), ensures that at least 60-70% of the total band-pass (YJ- or H-band) is completely free from OH airglow, e.g., Martini & DePoy (2000). For bright sources this is sufficient to extract good spectra, with the residual background between the sky lines removed by subtracting an average measurement of the sky. However, for faint sources – as in some of *MOONS* science cases – even between the OH sky lines an accurate sky-subtraction is needed. For this purpose, some sort of nodding-to-sky is required to reach an adequate signal-to-noise. This will be achieved by using a dedicated sky fiber per every object, positioned at a distance of few arcsec and performing cross-beam switching. A detailed description of the sky subtraction strategy envisaged for *MOONS* is presented by Rodrigues et al. (2012, these proceedings). Preliminary results obtained with FLAMES/GIRAFFE show that with cross-beam switching it is possible to subtract the sky background to better than 1% level (see Rodrigues et al. 2012).

# 4. THE MOONS BASELINE DESIGN

The baseline instrument concept for *MOONS* is a bench-mounted, multi-band spectrograph fed with fibres from a seeing-limited beam provided by the VLT. Each target is coupled to a fibre through a micro-lens in the focal plane, with the micro-lens and fibre assembly moved by a positioner. The full 25' diameter VLT Nasmyth focus is sub-divided into ~1000 sub-fields each patrolled by a single fibre, with the possibility of 2 neighboring sub-fields to partially overlap. The fibres will be divided between two identical tri-band spectrographs using dichroics to split the spectra for simultaneous observations. In this section we briefly highlight the main *MOONS* sub-systems.

## 4.1. Field corrector sub-system

In order to enable the exploitation of the full 25 arcminute field of view (FOV) of the VLT Nasmyth focus, the use of a field corrector is mandatory. The corrector is formed by two large lenses with approximately 110 mm of thickness in the axis and a diameter of 880 mm. The first lens is a plano-convex and the second is a symmetrical biconcave; this option allows minimizing manufacturing costs without influencing performances. To provide a glass internal transmission better than 95% for the full MOONS wavelength range (800 nm to 1800 nm), the selected material is Fused Silica. Compared to a no-corrector option, this corrector improves the image quality over the full FOV by a factor of 8, better than 0.1 arcsec (80% geometric energy for full FOV), the exit pupil is practically concentric to the field curvature and the field curvature is reduced to half (from a radius of 2090 mm to 4210 mm).

## 4.2. Fiber positioner sub-system

The fibers for science observations are deployed on the focal plane via a fiber positioner. Due to the different requirements on sky subtraction for bright and faint targets, the pick-off system must be able to allocate the 1000 fibers both independently – e.g. all on different targets – or in pairs to perform the cross-beam switching, in which each target has got its dedicated sky fiber at a distance of few arcsec. Another key requirement for the positioner is the reconfiguration time, which should be < 5 minutes in order to have acceptable overheads. For the Phase A study two possible implementations have been considered: a micro-mechanical pick-off system and a pick and place spine system.

### 4.2.1. Micro-mechanical pick-off system

In the micro-mechanical pick-off solution the idea is to cover the focal plane with modular fibre positioners each of which has a fixed patrol area. To cover the Nasmyth focal plane with ~1000 units, each positioner should have a physical size of ~30mm in diameter, which corresponds to ~1 arcmin on sky. In order to have a high allocation efficiency of the fibres on targets, some overlap between neighboring patrol fields is needed, with one fibre being able to patrol up to the centre of the neighboring cell.

In this concept design each pick-off unit has got two rotating arms. The fibre is positioned by rotating a second arm around the interconnection axis of the first arm, which rotates around a central axis in the centre of the patrol field, as shown in Figure 3. The positioner depicted uses two 8mm diameter geared Faulhaber stepper motors and speed reducers, with zero-backlash. There is a physical end-stop on the outer arm and datum micro-switches on both axes. A balance weight is positioned on the inner axis. It is possible to mount the pick offs from the back of the plate as the actuator assembles will pass through the large holes when the arms are retracted. This can facilitate assembly, maintenance and electronics packaging. The drive and position sensor electronics are located on the base of each positioner module, simplifying the electronics and cabling. Each module will be mounted on a flat facet, angled to the next to form a faceted curved surface to suit the focal plane curvature.

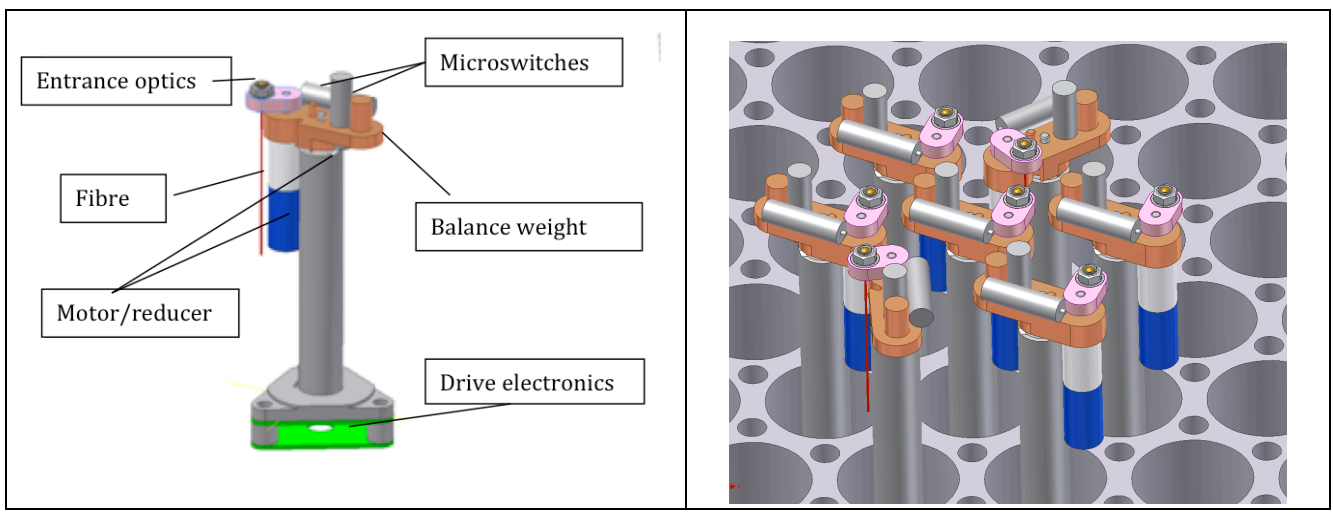

**Figure 3.** Micro-mechanical pick-off system.

### 4.2.2. Pick and place spine positioner

The second positioner design considered for the Phase A is a pick and place system. In this option each fibre button is supported by an individual spine passive support, and able to patrol an area, similar to the one for micro-mechanical positioner. Each fiber is re-positioned between configurations by several (7-10) radial arm pick and place positioner mechanisms similar to OzPoZ on FLAMES, as shown in the right panel of Figure 4. Multiple positioning mechanisms are required to meet the reconfiguration time requirements.

The spine support shown in the left panel of Figure 4 is comprised of three rods arranged to form a tripod. The rods pass through a split spherical bearing clamped in conical seats on the baseplate and clamp plate. Axial spring force on the clamp plate compresses the split ball onto the shaft fixing it in place. An actuator may be included in order to release the clamping force. The fibre angle is controlled by the gripper axis mounted on the positioner. As for the micro-mechanical system, some overlap between individual patrol fields is necessary to achieve high allocation efficiency. The spine design easily allows a 50% overlap in the patrol areas and close packing is limited only by the size of the fibre button and the gripper collett diameter.

The positioner mechanism is comprised of two radial arms mounted on rotary stages. The stages are envisaged to be direct drive with inbuilt absolute encoders. The gripper mechanism will provide a small amount of Z travel to enable an actuated collett to grip the fibre button, similar to OzPoZ design. This may include a camera to verify or fine tune the gripper position with respect to the fibre button. The positioning mechanisms are in front of the fibre pick off buttons and the fibres will be configured with the baseplate in a retracted position and then moved into place with an axial linear motion.

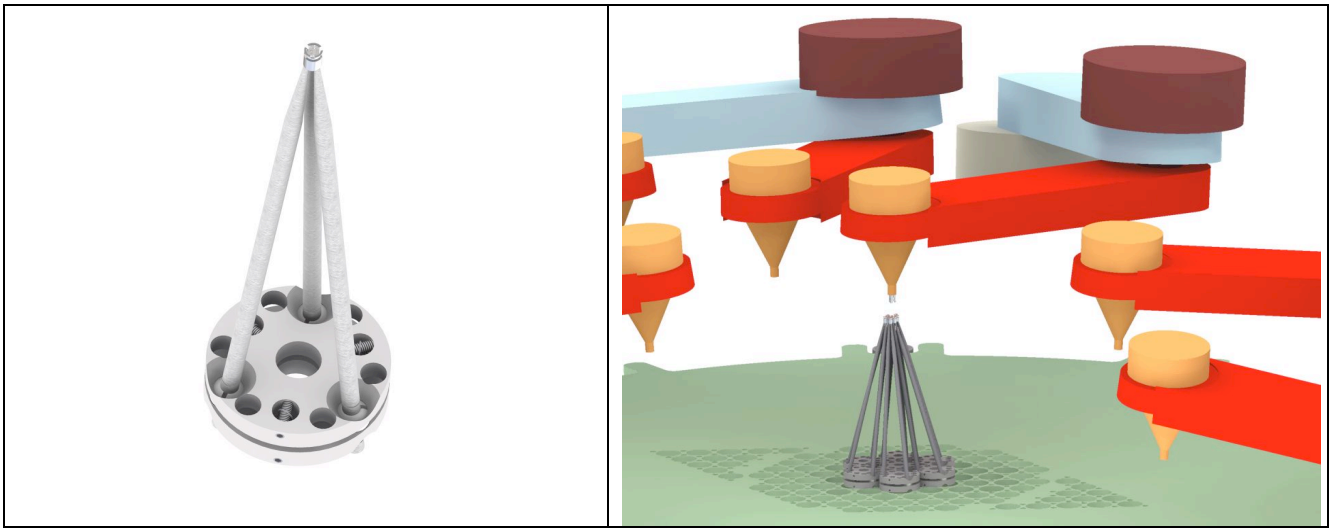

**Figure 4.** Pick and place spine positioner.

### 4.3. Fiber sub-system

At the entrance of the fibres, the optical aperture conversion from F/15 beam of the telescope to F/3.65 is realized by the coupling with a microlens. In fact, injecting at a fast F ratio allows the change between the entrance and output F ratio to be minimised. In fact, a system using an input of F/3.65 with an output of F/3.5 looses less than 2.5% in transmission. The input aperture on the sky of the fibre is 1-1.2 arcsec, which correspond to a physical core diameter of 140-170μm. In the phase A study only Polymicro fibres are investigated, since they are well known and are used in various astronomical instruments.

At the output end, a bundle of 10 to 15 fibres is grouped into sub-slits, which are then arranged to form the entrance slit to the spectrograph. The fibres inside the spectrograph enclosure are kept at cryogenic temperature, similarly to APOGEE (Brunner et al. 2010).

## 4.4. Spectrograph sub-system

A complete description of the *MOONS* spectrograph is presented by Oliva et al. (2012; these proceedings). The baseline design consists of two identical cryogenic spectrographs. Each of them collects the light from over 500 fibers and feeds, through dichroics, 3 spectrometers covering the "I" (0.79–0.94 μm), "YJ" (0.94–1.35 μm) and "H" (1.45–1.81 μm) bands, simultaneously (see Figure 5). The low resolution mode provides a complete spectrum with a resolving power ranging from R>4,000 in the YJ-band, to R>6,000 in the H-band and R>8,000 in the I-band. A higher resolution mode with R>20,000 is also included. It simultaneously covers two selected spectral regions within the YJ and H bands. The whole spectrometer is in a vacuum vessel cooled to cryogenic temperatures. Each spectrograph, utilizes two Hawaii-4RG-15 devices, from Teledyne Imaging (one for the YJ and one for H-band channel) and one 4k x 4k optical CCD from e2v technologies for the I-band channel.

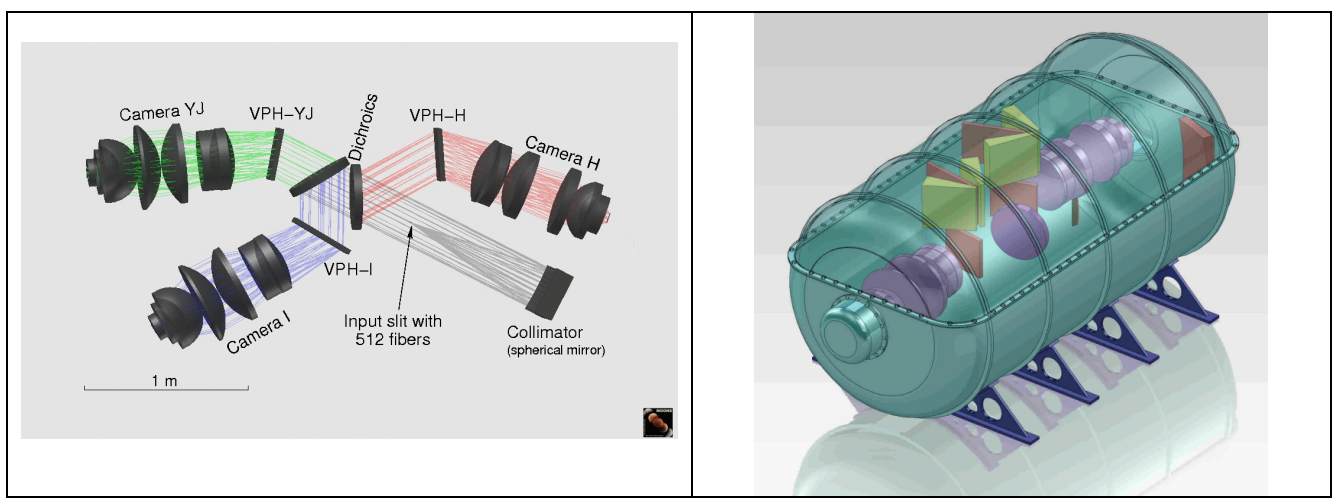

**Figure 5.** One of the two spectrographs envisaged for *MOONS,* for more details see Oliva et al. 2012, in these proceedings. The left panel shows the optical design, while the right panel shows the cryostat envelop.

## 5. SUMMARY

*MOONS* is a conceptual design study for a fibre-fed, multi-object spectrograph proposed as a new generation instrument for the VLT. The wealth of science that an efficient, highly-multiplexed, near-IR spectrograph can generate is undeniably vast and it has been a common aspiration for the VLT for long time.

This instrument will be extremely versatile and provide the ESO community with a world-leading work-horse capability able to serve a wide range of Galactic, Extragalactic and Cosmological studies, providing:

i) The necessary ground-based follow-up for the GAIA mission - the ESA cornerstone mission - to create the most accurate three-dimensional map of our Galaxy deriving fundamental insight into the origin and evolution of the Milky Way. Radial velocities will be obtained with a medium resolution via observations of the Calcium triplet and the high resolution mode (R~20,000) will provide detailed chemical abundances.

ii) Fundamental insight into galaxy formation and evolution by deep observations of galaxies at $z>1$, generating an SDSS-like survey at high redshift allowing an unprecedented study of the physical processes that shape galaxy evolution.

iii) The ideal instrument to follow up large area surveys/facilities in optical/near-IR (VISTA, UKIDSS, VST, Pan-STARRS, Dark Energy Survey, LSST, Euclid) and facilities operating at other wavelengths (ALMA, Herschel, eRosita, LOFAR, WISE, ASKAP).

To address such fundamental science goals *MOONS* will exploit the full 500 square arcmin field of view offered by the Nasmyth focus and be equipped with ~1000 fibers. Depending on the brightness of the targets, the fibres will be positioned all independently or in pairs allowing cross-beam switching for optimal sky subtraction. For the Phase A study we are investigating two possible implementations for the fiber positioner: a micro-mechanical pick-off system and a pick and place spine system. The light from the 1000 fibres will be divided between two identical tri-band spectrographs using dichroics to split the spectra for simultaneous observation of the full wavelength range 0.8μm - 1.8μm at a resolving power R>4,000. Three key narrower bands at ~0.85μm, ~1.2μm and ~1.6μm will also be covered simultaneously at higher resolving power to determine radial velocities and detailed chemical abundances (R≥7,000 around CaII triplet and R≥20,000 in the J and H bands).